\documentclass[pra,bibnotes,twocolumn]{revtex4}
\usepackage{graphicx}
\begin{document}
\draft
\def\ds{\displaystyle}
\title{ Stable topological edge states in a non-Hermitian four-band model }
\author{C. Yuce}
\address{Department of Physics, Anadolu University, Turkey }
\address{Department of Physics, Eskisehir Technical University, Turkey }
\email{cyuce@anadolu.edu.tr}
\date{\today}
\begin{abstract}
We introduce a one dimensional non-Hermitian four band tight binding lattice system. We find stable topological edge states protected by particle-hole and parity-time symmetries. We show that topological phase appears in the system. We discuss that the system can also be used as a topological laser if the gain and loss positions in a unit cell are judiciously arranged. 
\end{abstract}
\maketitle

 \section{Introduction}
 
In the last decade, tremendous progress has been made in our understanding of topological phases. Many theoretical models  that are difficult to realize in condensed matter physics using electrons can be experimentally realized in photonics using coupled waveguides and optical lattices. In recent years, the concept of the topological phase has been extended to non-Hermitian systems, which presents new physics inaccessible in Hermitian systems. This new subfield attracts researchers studying topological photonics and $\mathcal{PT}$ symmetric systems, where $\mathcal{P}$ and $\mathcal{T}$ are parity and time reversal operators, respectively. The $\ds{\mathcal{PT}}$ symmetry in photonics is implemented by replacing the Schrodinger equation into the Helmholtz equation. In this way, a $\ds{\mathcal{PT}}$ symmetric optical system can be realized if the real and imaginary parts of the refractive index are symmetric and antisymmetric, respectively. The antisymmetric character of the imaginary part of the refractive index implies that the system has symmetrically located gain and loss. The first papers in non-Hermitian topological systems found no stable topological insulating phase \cite{PTop3,ekl56,PTop4,PTop1}. Instead they found either decaying or growing topological modes. In 2015, it was theoretically predicted that stable topological phase is compatible in a non-Hermitian Aubry-Andre model  \cite{cem0001}. This prediction was followed by an elegant experiment using waveguides \cite{sondeney1}. In that experiment, stable topological zero energy modes were observed through fluorescence microscopy in a photonic lattice which consists of waveguides with staggered hopping amplitudes. So far, no stable topological phase has been to shown to exist in a two dimensional non-Hermitian system. Almost all papers in this subfield deals with one dimensional problems \cite{sbt1,sbt3,sbt6,sbt11,sbt8,sbt2,sbt500,cyucepra}. Complex extensions of the celebrated Su-Schrieffer-Heeger (SSH) model was shown to support stable topological zero energy modes \cite{sbt4,cmyc}. Note that topological phase is not restricted to topological insulators. Topological superconductors with gain and loss are also emerging field and Majarona modes have been studied in the context of non-Hermitian systems \cite{sbt12,sbt16,sbt17,sbt7,sbt18}. Topological phase may also appear in time-periodic systems. This is known as Floquet topological phase \cite{flotop1,flotop11} and was studied in non-Hermitian systems \cite{sbt10,sbt15}. \\
Despite the progress of non-Hermitian topological photonics, the topic is still in its infancy. We need more theoretical models in $1D$ in addition to the SSH model and keep looking for stable topological phase in higher dimensions. In \cite{sbt3}, topological edge states in a non-Hermitian trimerized tight binding lattice have recently been studied but the considered model has complex energy eigenvalues. In this paper, we study another one-dimensional non-Hermitian model that supports stable topological edge states. We extend a non-Hermitian generalization of the four-band model \cite{klaidescop} and show that  the extended non-Hermitian system has stable topological phase protected by particle- hole symmetry. Furthermore, depending on the gain and loss arrangement, it can also be an ideal system to realize 1D topological laser. It is generally believed that non-Hermitian photonics topological insulators can one day be used in technology. Recently, topological lasers have attracted great attention \cite{topinsu,topinsu278}. Finally we note that our model can be experimentally realized with current technology using waveguides or microrings.

 \section{Formalism}

The most extensively studied one dimensional tight-binding topological insulating model is the Su-Schrieffer-Heeger (SSH) model, which is a two-band system since each unit cell has $\ds{2}$ different hopping amplitudes. Here we consider a tight-binding lattice with four different hopping amplitudes in each unit cell. This four-band model is a special case of the $\ds{1D}$ off-diagonal Aubry-Andre model \cite{cmyc}. Suppose that the chain consist of $N/4$ unit cells, each unit cell hosting $\ds{4}$ sites, so the total number of sites is $N$. In addition, suppose gain and loss with site-dependent non-Hermitian degree $\ds{\gamma_n}$ are introduced into the system. Note that $\ds{\gamma_{n+4}=\gamma_n}$ since the unit cell repeats itself at every 4 sites. If the non-Hermitian degree is positive (negative), then the corresponding site has gain (loss). We require that the non-Hermitian impurities are arranged in a balanced way in each unit cell $\ds{\gamma_1+\gamma_2+\gamma_3+\gamma_4=0}$. Adding gain and loss to one site in the unit cell implies that we insert positive and negative values of  $\gamma$ on the diagonal of the Hamiltonian, respectively. The complex Hamiltonian $H_C$ reads
\begin{eqnarray}\label{mcabjs4}
H_C=&-&t\sum_{n=1}^{N-1}\left(1+\lambda \cos{(\frac{\pi}{2} n+\Phi)}\right)  a^{\dagger}_{n} a_{n+1}+H.c.\nonumber\\
&+&i\sum_{n=1}^{N}\ \gamma_n ~ a^{\dagger}_n a_n 
\end{eqnarray}
where $\ds{t}$ is unmodulated tunneling amplitude, $-1<\lambda<1$ is the modulation strength, the modulation phase $\ds{\Phi}$ is an additional degree of freedom, $\ds{a^{\dagger}_n}$ and $\ds{a_n}$ denote the creation and annihilation operators on site $\ds{n}$, respectively. For the sake of simplicity, we take $\ds{t=1}$. This system can be experimentally realized in a photonic lattice using waveguides or microrings. For example, four different tunneling amplitudes in the unit cell can be realized by adjusting the distance between the neighboring waveguides. If the waveguides are provided gain and loss in such a way that the net gain and loss in each unit cell is zero, then the above Hamiltonian can be realized. Note also that our system can be mapped directly onto a system with only loss as explained in \cite{sondeney1}.\\
Our first aim is to rewrite the above Hamiltonian in momentum space. We do this for a special value of the modulation phase for which the non-Hermitian Hamiltonian has particle-hole symmetry. As we shall see below, the system is particle-hole symmetric if $\ds{\Phi=0}$. Therefore, the hopping amplitudes in a unit cell are given by $\ds{\{1,1-\lambda,1,1+\lambda\}}$. The corresponding non-Hermitian Hamiltonian in momentum space can now be found using $\ds{
H_C=\sum_{k=0}^{\infty}    (a_{1,k}^{\dagger},a_{2,k}^{\dagger},a_{3,k}^{\dagger},a_{4,k}^{\dagger}  ) ~ \mathcal{H_C}(k)~ \left(\begin{array}{cc}a_{1,k}  \\a_{2,k} \\a_{3,k} \\a_{4,k}\end{array}     \right)}$ with the Fourier transformation $\ds{a_{j,k}=1/\sqrt{N/4} ~\sum_{k} e^{ikm}a_{j,m}  }$, where $\ds{{a}_{j,m}}$ and $\ds{{a}^\dagger_{j,m}}$ are the annihilation and creation operators localized at cell number $m=1,2,...,N/4$ for $j=1,2,3,4$ in each unit cell, respectively. Note that each unit cell includes 4 sites per cell, so we rewrite the annihilation operator on site $\ds{n}$ as $\ds{ a_n=(a_{1,m} ,a_{2,m} ,a_{3,m} ,a_{4,m}   ) }$. Let us start with the Hermitian part $\ds{\mathcal{H} (k)}$ of our Hamiltonian $\ds{\mathcal{H_C} (k)}$ at $\ds{\Phi=0}$
\begin{eqnarray}\label{mtdlpo2}
\mathcal{H} (k)=\mathcal{I} \otimes   \sigma_x+\frac{\lambda^+\sin k}{2}(\tau_y \otimes   \sigma_x+\tau_x \otimes   \sigma_y)+\nonumber\\
~~~\frac{\lambda^- - \lambda^+\cos k}{2} \tau_y \otimes   \sigma_y+\frac{\lambda^- + \lambda^+\cos k}{2}\tau_x \otimes   \sigma_x
\end{eqnarray}
where $\ds{\lambda^-=1-\lambda}$ and $\ds{\lambda^+=1+\lambda}$, $\mathcal{I}$ is the identity matrix, $\ds{\sigma_i}$, $\ds{\tau_i}$ are Pauli matrices and $k$ is the crystal momentum defined in the first Brillouin zone, $-\pi<k<\pi$. \\
To discuss symmetry properties of this Hamiltonian, let us write the definitions of the antiunitary particle-hole operator $\ds{ \mathcal{C} }$ and the unitary parity operator $\ds{ \mathcal{P} }$
\begin{eqnarray}\label{mwpoasj}
\mathcal{C} ~\mathcal{H_C} (-k)~\mathcal{C}^{-1}&=&-\mathcal{H_C} (k);  \nonumber\\
\mathcal{P} ~\mathcal{H_C} (-k)~\mathcal{P}^{-1}&=&~~\mathcal{H_C}(k).
\end{eqnarray}
\begin{figure}[t]\label{2tg0}
\includegraphics[width=8.5cm]{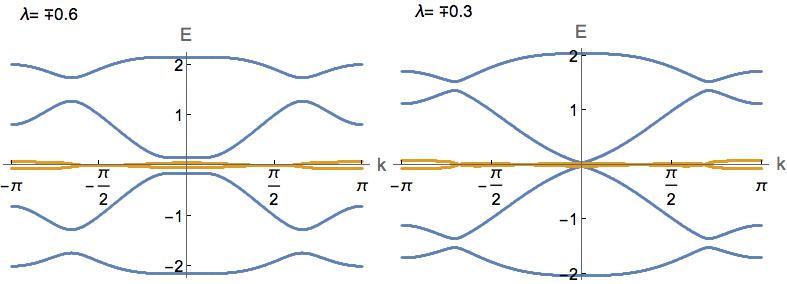}
\includegraphics[width=8.5cm]{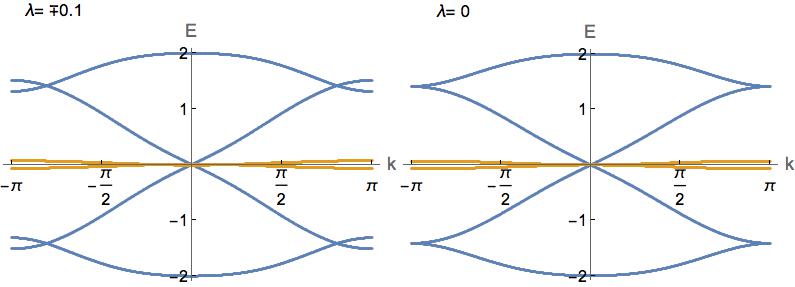}
\caption{ The energy spectra for the Hamiltonian (\ref{myytrudk2}) at $\ds{\lambda=\mp 0.6}$ and $\ds{\lambda=\mp 0.3}$ in the upper panel and $\ds{\lambda=\mp 0.1}$ and $\ds{\lambda=0}$ in the lower panel. The real and imaginary parts of the energy spectrum at $ \ds{\gamma_0=0.2}$ are depicted in blue and orange colors, respectively. The spectrum is complex valued for all values of $\lambda$ since the system has no $\ds{\mathcal{PT}}$ symmetry. }
\end{figure}
\begin{figure}[t]\label{2678ik0}
\includegraphics[width=8.5cm]{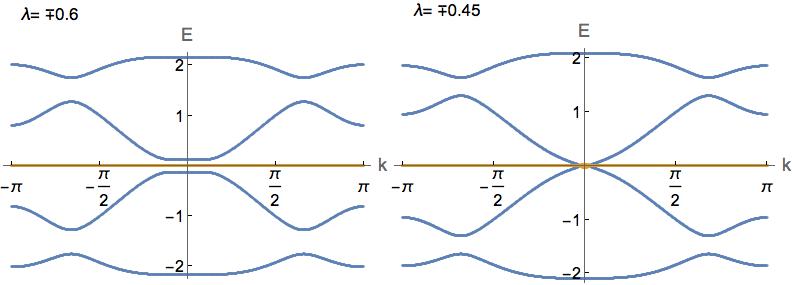}
\includegraphics[width=8.5cm]{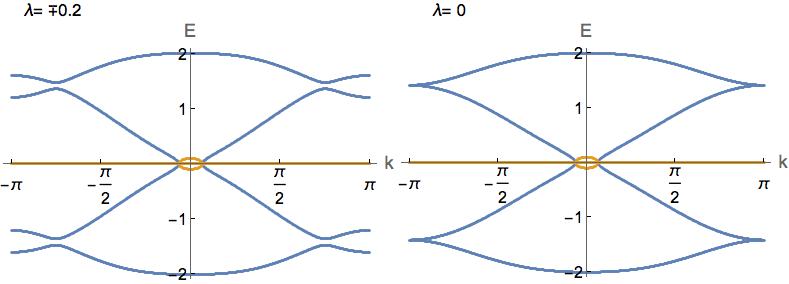}
\caption{ The energy spectra for the Hamiltonian  (\ref{mythfncz2}) at $\ds{\lambda=\mp 0.6}$ and $\ds{\lambda=\mp 0.45}$ in the upper panel and $\ds{\lambda=\mp 0.2}$ and $\ds{\lambda=0}$ in the lower panel. The real and imaginary parts of the energy spectrum at $ \ds{\gamma_0=0.2}$ are depicted in blue and orange colors, respectively. The $\ds{\mathcal{PT}}$ symmetriy is spontaneously broken for the middle two bands and exceptional points occur at a non-zero value of $\lambda$. The top and the bottom bands retains the $\ds{\mathcal{PT}}$ symmetry and the two bands are real valued as long as the non-Hermitian degree doesn't exceed the critical value. }
\end{figure}
Our system has various symmetry protected topological phases depending on the form of $\ds{\gamma_n}$. In the following, we study three different choices of $\gamma_n$. These are given by $\ds{\{+i\gamma_0,-i\gamma_0,-i\gamma_0,+i\gamma_0\}}$, $\ds{\{+i\gamma_0,+i\gamma_0,-i\gamma_0,-i\gamma_0\}}$ and $\ds{\{+i\gamma_0,-i\gamma_0,+i\gamma_0,-i\gamma_0\}}$, where $\ds{\gamma_0}$ is a real-valued constant. We note that the spectrum remains the same under $\ds{\gamma_0\rightarrow-\gamma_0}$. Below, we will firstly find energy spectra and discuss topological phase transitions for periodical lattices and then perform numerical calculation for finite lattices with open edges. To see topological phase transition in the periodical lattice, we tune $\ds{\lambda}$ from a positive value to a negative value. We numerically see that $\ds{E(k)\rightarrow  {E}(k)}$ for the first and the third choices of $\gamma_n$ and $\ds{E(k)\rightarrow  {E^{\star}}(k)}$ for the second choice of $\gamma_n$ under the sign of the modulation strength $\ds{\lambda\rightarrow -\lambda}$. Furthermore the energy bands are symmetric around zero energy. Therefore there are two immediate results. First, $\ds{E}$ versus $\ds{k}$ plots are the same for $\ds{\mp\lambda}$. Second, band gap closing (for the real part of the spectrum) occurs either at a single point $\ds{\lambda=0}$ or in a finite symmetric interval of $\ds{\lambda}$ around zero $(-\lambda_0,\lambda_0)$. In the latter one, exceptional points occurs and then complex energy eigenvalues appear. There is one more distinctive feature of the non-Hermitian character. There are 3 band gaps in the four-band system. All these 3 band gaps close simultaneously at $\lambda=0$ in the Hermitian system. However, this is not the case in the non-Hermitian system. Instead, band gap closings occur sequentially. In other words, decreasing $\ds{\lambda}$ closes one band gap at a specific value of $\lambda$ while the other bands are still gapped. To close the two other bands, we need to decrease $\ds{\lambda}$ more. As it is well known, band gap closing is a signature of the topological phase transition. Therefore, we conclude that topological phase transition occur one after the other in the same system.\\
Suppose first that gain and loss are arranged in each sublattice as follows: $\ds{\{   +i\gamma_0,-i\gamma_0,-i\gamma_0,+i\gamma_0 \}}$. Therefore the non-Hermitian Hamiltonian is given by
\begin{eqnarray}\label{myytrudk2}
\mathcal{H_C} (k)=i \gamma_0 ~ \tau_z \otimes   \sigma_z+\mathcal{H} (k)
\end{eqnarray}
where the Hermitian part $\ds{\mathcal{H} (k)}$ is given by (\ref{mtdlpo2}). One can see that this non-Hermitian Hamiltonian has both parity $\ds{\mathcal{P}=\tau_x  \otimes  \sigma_x}$ and particle-hole symmetry $\ds{\mathcal{C}=\mathcal{I} \otimes  \sigma_z~\mathcal{K}}$, where $\ds{\mathcal{K}}$ is the complex conjugation. Let us study energy spectrum for this system for $\ds{\gamma_0=0.2}$. There are four bands as can be seen from the Fig-1. In the figure, the real (complex) part of the energy eigenvalues is depicted in blue (orange) color. Unfortunately, the system has complex energy at every value of $\ds{\lambda}$. This is because of the fact that the system has parity symmetry but not $\ds{\mathcal{PT}}$ symmetry. For large values of $\ds{|\lambda|}$, the real part of the spectrum is gapped. If we decrease $\ds{|\lambda|}$, then the band gap of the two bands in the middle gets narrower. The band gap closing at the center of the Brillouin zone occurs at $\ds{\lambda = 0.3}$ and is kept closed in the interval $\ds{-0.3<\lambda<-0.3}$. Note that the top and bottom bands are still gapped at $\ds{|\lambda|=0.3}$ and their band gaps close at a smaller value of $|\lambda|$ around the edges of the Brillouin zone. At this special value of $\lambda$, the system becomes fully gapless. Decreasing $\lambda$ shifts the band touching points and the top and bottom bands touch their neighboring bands at exactly the edges of the Brillouin zone when $\lambda=0$. As a result, we expect either growing or decaying topological edge states because of the appearance of complex energy eigenvalues during topological phase transition. Furthermore, we expect that bulk states have complex energy eigenvalues as well because of the absence of parity-time symmetry. To this end, we note that increasing $\gamma_0$ increases the absolute values of the imaginary parts of the energy eigenvalues. For very large values of $\gamma_0$, the spectrum becomes purely imaginary in the whole Brillouin zone.\\
\begin{figure}[t]\label{2tg0}
\includegraphics[width=8.5cm]{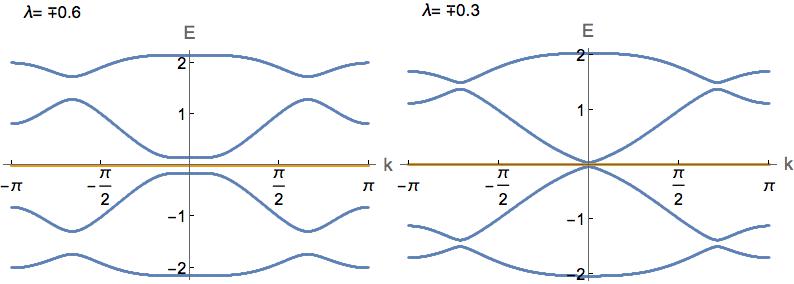}
\includegraphics[width=8.5cm]{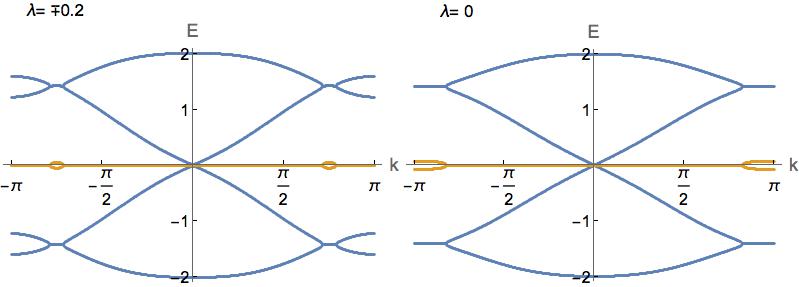}
\caption{ The energy spectra for the Hamiltonian  (\ref{mrtgvkk2}) at $\ds{\lambda=\mp 0.6}$ and $\ds{\lambda=\mp 0.3}$ in the upper panel and $\ds{\lambda=\mp 0.2}$ and $\ds{\lambda=0}$ in the lower panel. The real and imaginary parts of the energy spectrum at $ \ds{\gamma_0=0.2}$ are depicted in blue and orange colors, respectively. As opposed to the case in the Fig.2, the $\ds{\mathcal{PT}}$ symmetry is spontaneously broken not for the middle two bands but for the top and the bottom bands.}
\end{figure}
Let us now study the second choice with $\ds{\{+i\gamma_0,+i\gamma_0,-i\gamma_0,-i\gamma_0\}}$, where the first two sites in each unit cell have gains while the other two sites have losses. The non-Hermitian Hamiltonian in momentum space is then given by
\begin{eqnarray}\label{mythfncz2}
\mathcal{H_C} (k)=i \gamma_0~\mathcal{I} \otimes   \sigma_z+\mathcal{H} (k)
\end{eqnarray}
This Hamiltonian has particle-hole symmetry $\ds{\mathcal{C}=\mathcal{I} \otimes  \sigma_z~\mathcal{K}}$ but not parity symmetry. Instead it has parity-time symmetry $\ds{\mathcal{PT}=\tau_x\otimes  \sigma_x~\mathcal{K}}$. Note that $\ds{\mathcal{PT} ~\mathcal{H} (k)~(\mathcal{PT})^{-1}=~~\mathcal{H} (k)}$. This case is interesting since the top and bottom bands are always real-valued as long as $\gamma_0$ doesn't exceed a critical value and they touch their neighboring bands at the Brillouin zone when $\ds{\lambda=0}$ as can be seen from the Fig-2. These band gap closing points are not exceptional points and they open again if we tune $\lambda$ to negative values. This shows us that the $\ds{\mathcal{PT}}$ symmetry is not broken for the top and bottom bands during the topological phase transition. However, the picture changes considerably for the two middle bands. The $\ds{\mathcal{PT}}$ symmetry is broken for these two middle bands and the corresponding band gap closing occur at a non-zero value of $\lambda$. At this particular value of $\lambda$, exceptional points where the eigenstates are no longer the eigenstates of the $\ds{\mathcal{PT}}$ operator appear. Then these two bands have complex energy if we further decrease $\lambda$. In other words, complex energy eigenvalues appear as middle band gap closes and reopens. Note that the spectrum is not complex for the whole Brillouin zone but in a small interval around the Brillouin zone center. As a result, we say that $\ds{\mathcal{PT} }$ symmetry is spontaneously broken for the two middle bands while it is retained for the top and bottom bands during the topological phase transition. \\
For the final case, we consider the case with alternating gain and loss in the system, i..e., $\ds{\{+i\gamma_0,-i\gamma_0,+i\gamma_0,-i\gamma_0\}}$. The corresponding non-Hermitian Hamiltonian reads
\begin{eqnarray}\label{mrtgvkk2}
\mathcal{H_C} (k)=i \gamma_0~  \tau_z \otimes  \mathcal{I} +\mathcal{H} (k)
\end{eqnarray}
\begin{figure}[t]\label{2tg0}
\includegraphics[width=9cm]{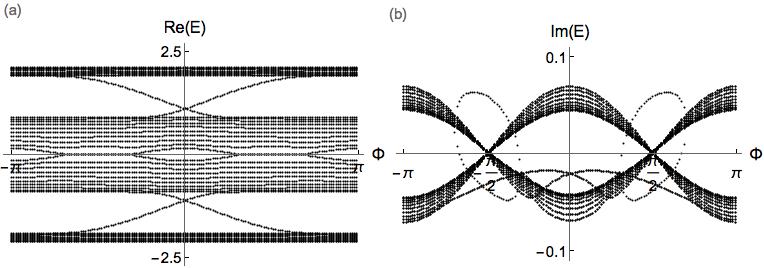}
\includegraphics[width=9cm]{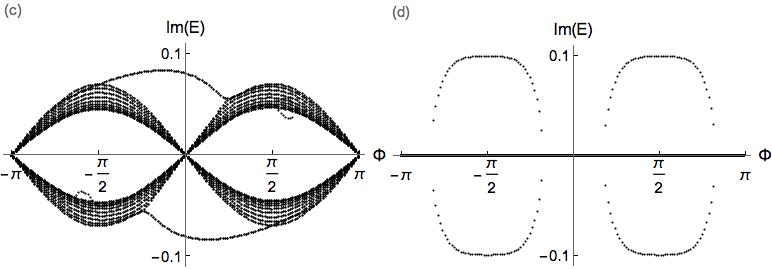}
\caption{ The energy spectra as a function of $\ds{\Phi}$ at $ \ds{\gamma_0=0.1}$ and $\lambda=0.5$. The first figure (a) plots the real part of the energy spectrum for the Hamiltonian (\ref{mrtgvkk2}). Note that the real part of the energy eigenvalues are almost the same for two other complex Hamiltonians. The figures (b), (c) and (d) depict the imaginary parts of the energy eigenvalues for the Hamiltonians (\ref{myytrudk2}, \ref{mythfncz2}) and (\ref{mrtgvkk2}), respectively. All three non-Hermitian Hamiltonians has particle-hole symmetries at $\ds{\Phi=0}$. However, only the system in (b) has no $\ds{\mathcal{PT}}$ symmetry, which leads to complex energy eigenvalues. The $\ds{\mathcal{PT}}$ symmetry is spontaneously broken in (c) when the system has open edges. As a result, topological lasing can be studied for the Hamiltonian (\ref{mythfncz2}) since all the bulk states have real valued energy while topological edge states have non-vanishing imaginary part at $\Phi=0$. Finally, the $\ds{\mathcal{PT}}$ symmetry exist for both periodic and lattice form for the Hamiltonian (\ref{mrtgvkk2}) at $\ds{\Phi=0}$, so stable topological edge states can be realized. }
\end{figure}
The particle-hole symmetry exists in this Hamiltonian, too: $\ds{\mathcal{C}=\mathcal{I} \otimes  \sigma_z~\mathcal{K}}$. It is also $\ds{\mathcal{PT}}$ symmetric: $\ds{\mathcal{PT}=\tau_x\otimes  \sigma_x~\mathcal{K}}$. We plot the corresponding energy spectrum in the Fig-3. This case is the opposite of the previous case in the sense that $\ds{\mathcal{PT} }$ symmetry is spontaneously broken during the topological phase transition not for the middle two bands but for the top and bottom bands. Another difference here is that exceptional points occur at around the edges of the Brillouin zone, while they occur at the Brillouin zone in the previous case. \\
So far, we have studied the energy spectra for periodical systems. Below we perform numerical computation to find spectra for finite lattices with $\ds{N=40}$ for the above three Hamiltonians. We plot energy spectra as a function of $\Phi$ to see topological insulating states more clearly. Note that only the case with $\ds{\Phi=0}$ is important since the Hamiltonian (\ref{mwpoasj}) has particle-hole symmetry at this particular value of $\ds{\Phi}$. We numerically see that the real parts of the energy spectrum are almost the same for these three cases. The Fig-4 (a) shows the spectrum as a function of $\ds{\Phi}$. Inside the band gap between the top two bands (or between the bottom two bands), the topological states can be observed. The edge modes are found to be robust against weak coupling constant disorder, which can be introduced in our system by having randomized weak modulation strength in the lattice, i.e., $\ds{\lambda\rightarrow \lambda+\delta\lambda_n}$, where $\ds{|\delta\lambda_n|<<\lambda}$ is site-dependent real-valued random set of constants. The robustness against such a symmetry protected disorder is a direct result of the topological nature of the edge states. A question arises. Are the topological edge states stable or do they grow (decay) in time due to the complex nature of the spectrum? If we look at the imaginary parts of the spectrum at $\Phi=0$ in the Fig-4 (b) to (d), we see that there are many states with complex energy eigenvalues in (b) while there are two such states in (c) and no such state in (d). This implies that only the non-Hermitian Hamiltonian (\ref{mrtgvkk2}) has stable topological edge states. Furthermore, both edge and bulk states have complex energy eigenvalues for the Hamiltonian (\ref{myytrudk2}) while only edge states have complex energy eigenvalues for the Hamiltonian (\ref{mythfncz2}). The latter one can be used to realize a topological laser since the bulk states are stable but the robust topological edge state can grow in time. Note that the left edge has gain impurity while the right edge has loss impurity when $N=40$. Therefore only the left edge can be used for a topological laser. If $N=42$ instead, then both edges terminate with gain materials. In other words both edges can be used for a topological laser. We refer the reader the reference \cite{topinsu278} for the working principle of the topological laser.\\
One may wonder why the Hamiltonian (\ref{mythfncz2}) admits growing/decaying topological edge states while the Hamiltonian (\ref{mrtgvkk2}) has stable topological edge states although both of them are $\ds{\mathcal{PT}}$ symmetric. This can be understood as follows. The two systems are $\ds{\mathcal{PT}}$ symmetric so bulk states for the two systems have real valued energy eigenvalues. It is well known that topological edge states are well localized around the edges. The first two sites at the edges of the finite chain have gains and the next two sites have losses in (\ref{mythfncz2}) while gain and loss changes in an alternating way along the lattice in (\ref{mrtgvkk2}). Therefore, topological edge states see gain materials dominantly for the former case, which in turn makes the topological edge states complex valued. Finally, let us discuss topological classification of our system. According to the standard classification of topological insulators, our system has particle-hole  symmetry so it is in class $D$. We note that the Hermitian part (\ref{mtdlpo2}) is in class $AIII$ since it has Chiral symmetry. Therefore the Hermitian part has $Z$ topological phase.\\
To sum up, we have studied a one dimensional four band model in a non-Hermitian system. We have analyzed the reality of the energy eigenvalues of the topological edge states for three different choices of the gain and loss arrangements. We have shown that symmetry protected topological phase with real spectrum appears in our system. We have discussed that the system has topological phase and topological edge states are protected by particle-hole and parity-time symmetries. Our model can be used to obtain stable topological edge states and one dimensional topological laser.

\end{document}